# Photocurrent generation at ABA/ABC lateral junction in tri-layer graphene photodetector

Minjung Kim [a], Seon-Myeong Choi [b], Ho Ang Yoon [c], Sun Keun Choi [a], Jae-Ung Lee [a], Jungcheol Kim [a], Sang Wook Lee [c], Young-Woo Son [b], and Hyeonsik Cheong [a,*]

[a] Department of Physics, Sogang University, 35 Baekbeom-ro, Mapo-gu, Seoul 121-742, Korea

[b] School of Computational Sciences, Korea Institute for Advanced Study, 85 Hoegi-ro, Dongdaemun-gu, Seoul 130-722, Korea

[c] Division of Quantum Phases and Devices, School of physics, Konkuk University, 120 Neungdong-ro, Gwangjin-gu, Seoul 143-701, Korea

Metal-graphene-metal photodetectors utilize photocurrent generated near the graphene/metal junctions and have many advantages including high speed and broad-band operation. Here, we report on photocurrent generation at ABA/ABC stacking domain junctions in tri-layer graphene with a responsivity of 0.18 A/W. Unlike usual metal-graphene-metal devices, the photocurrent is generated in the middle of the graphene channel, not confined to the vicinity of the metal electrodes. The magnitude and the direction of the photocurrent depend on the back-gate bias. Theoretical calculations show that there is a built-in band offset between the two stacking domains, and the dominant mechanism of the photocurrent is the photo-thermoelectric effect due to the Seebeck coefficient difference.

*Corresponding author. Tel: 82 2 705-8434. E-mail: hcheong@sogang.ac.kr (Hyeonsik Cheong)

# 1. Introduction

Photocurrent generation in graphene-based devices has attracted much interest since the broadband absorption [1,2] and high mobility [3,4] of graphene are advantageous for fast and broadband photodetectors [5–24]. However, in metal-graphene-metal photodetectors, the photocurrent is largely confined to the area immediately next to the metal electrodes. If the same metal is used for the electrodes, the photocurrent from the two electrodes cancels each other. Using different metals for the two electrodes can solve this problem, but fine-tuning of the band-alignment using the back-gate is required [11,15]. Dual gating [13,14] has also been used to create a lateral pn junction in the graphene channel, but a complicated fabrication process is required.

In general, photocurrent is generated by several different mechanisms [17], including photovoltaic [5–9,11,15], photo-thermoelectric [10,13,14], and bolometric [16] effects. The Fermi energy difference between two semiconductors results in a potential difference when a junction is formed. This potential difference separates photogenerated electrons and holes to result in a photocurrent (photovoltaic effect). A difference in the Seebeck coefficient can also generate photocurrent when light absorption causes a temperature gradient in the device (photo-thermoelectric effect). These effects can occur without external bias, but bolometric effect which is based on the conductance difference induced by heating requires an external bias.

In few-layer graphene obtained by mechanical exfoliation from natural graphite, two stacking orders, ABA or Bernal stacking and ABC or rhombohedral stacking, are found [25–34]. The different stacking sequences result in different physical properties due to the interlayer interactions. Therefore, a junction between ABA- and ABC-stacked graphene domains with the same thickness would behave like a heterojunction between two different semiconductors, and so photocurrent generation can be expected. Here, we report on observation of photocurrent in ABA/ABC junctions in tri-layer graphene (TLG) and present theoretical analysis for the photocurrent mechanism. In metal-graphene-metal devices, the photovoltaic effect dominates due to the large Fermi energy difference at the metal/graphene junctions [5–9,11,15]. On the other hand, at bi-/single-layer graphene (BLG/SLG) junctions, the photocurrent is predominantly due to the photo-thermoelectric effect [10]. In ABA/ABC junctions, the two effects are expected to coexist, and only careful comparison of experimental data with theoretical analysis can elucidate the photocurrent mechanism.

## 2. Experimental

2.1 Sample preparation

Graphene samples were mechanically exfoliated on heavily p-type silicon substrates covered with 300-nm thick silicon dioxide. TLG samples with both ABA and ABC stacking domains were found by measuring Raman images of several TLG samples [28,29,34]. BLG/SLG samples were found by optical microscopy and then confirmed by Raman imaging [35]. Palladium electrodes (35nm) capped with gold (35nm) were deposited by e-beam lithography and thermal evaporation. The substrate was used as the back-gate. The charge neutrality point (CNP) was found by measuring the resistance of the device as a function of the back-gate bias voltage (Fig. S1, Supplementary data). It should be noted that if there is a band offset as discussed below, the Fermi level may not coincide with Dirac points of the ABA or ABC domains at CNP.

2.2 Photocurrent and Raman measurement

All measurements were carried out in a vacuum of $10^{-5}$ Torr by using a micro vacuum chamber (Oxford Microstat He2) in order to minimize shift of the CNP during the measurements. Photocurrent images were obtained by raster-scanning a focused laser beam of 514.5-nm wavelength, chopped at 410 Hz. A long-working-distance objective (40×, N.A. 0.6) with a correction ring was used to focus the laser beam. The Beam spot size was about 1 μm. The photocurrent was measured by using the lock-in technique for good signal-to-noise ratios. During the raster scanning, a Raman spectrum was measured at each point in order to obtain a Raman image simultaneously. The Raman signal was dispersed by a spectrometer with a focal length of 550 mm (Horiba Triax 550) with a grating with 1200 grooves/mm and detected using a charge-coupled device (CCD). The laser power was 280 μW for ABA/ABC photodetectors and 480 μW for BLG/SLG devices. For confirmation of ABA and ABC stacking domains, we measured high-resolution Raman spectra from representative positions in each domain and observed Raman N, M, and 2D bands (Figs. 1b and c). For these measurements, a grating with 2400 grooves/mm was used to obtain a spectral resolution of ~0.36 cm$^{-1}$.

2.3 Theoretical calculations

To obtain the Seebeck coefficient and the work function, the electronic band structures were calculated using tight-binding methods with the hopping parameters obtained through ab initio density functional theory (DFT) and many-body quasiparticle calculations within the GW approximation [36]. The Seebeck coefficient calculation is based on the semi-classical Boltzmann transport theory with rigid band and constant relaxation time approximations [37–42]. The behaviors of the Seebeck coefficients are comparable to the results reported by Wierzbowska *et al.* although the graphs look somewhat different due to different scaling of the x-axis [43].

## 3. Results and discussion

Fig. 1a shows a schematic of the ABA/ABC photodetectors. We measured three similar devices, and the data from one of them are mainly analyzed here. For comparison, we also studied lateral junctions between SLG and BLG. Figs. S2a and b (Supplementary data) show optical images of the ABA/ABC and the BLG/SLG photodetectors, respectively. Figs. 1b and c show Raman modes that are used to identify ABA- and ABC-stacked domains in TLG [28,29,32,34]. The black spectrum is for ABA stacking and the red spectrum for ABC stacking. Because of different Raman 2D band shapes, the stacking domains can be easily identified by fitting the spectra with a *single*-Lorentzian function and imaging the peak position or the width [28,29].

Because the stacking domains cannot be distinguished in optical microscope images, we simultaneously measured the photocurrent and Raman spectra of the devices in order to correlate the photocurrent with the exact location of the ABA/ABC junction in the sample. This has the added benefit of removing the uncertainty due to drift of the system because the spatial position of the sample is monitored by Raman spectroscopy in situ. Fig. 2 compares the photocurrent images with corresponding Raman images for the 3 devices measured. They show that the photocurrent is mostly generated in the graphene channel at the ABA/ABC junctions. In order to elucidate the mechanism of the photocurrent at the ABA/ABC heterojunction, the photocurrent was measured as a function of the back-gate voltage (Fig. 3a). As the back-gate voltage is changed from positive to negative with respect to the CNP, the photocurrent direction is reversed (Fig. 3b). For comparison, similar measurements were carried out on a BLG/SLG lateral heterojunction photodetector (Fig. S3, Supplementary data). These results are consistent with the measurements of Xu *et al.* [10].

Here, we consider two possible mechanisms for photocurrent without an external bias. Suppose that both ABA and ABC TLG are at CNP initially. When the gate voltage is applied, the numbers of induced carriers in the two domains are different due to different density of states. This difference in the carrier density would induce difference in the Fermi energy. However, because the Fermi energy should be the same across the junction at equilibrium, the bands should actually realign with respect to each other so as to keep the Fermi energy uniform across the device, resulting in relative shifts of the Dirac points in the two regions (Fig. S4a, Supplementary data). This difference in the Dirac point energy would create a built-in potential difference which could drive the photocarriers to generate photocurrent. This is the photovoltaic effect. If the temperature of the junction is raised due to laser-induced heating with respect to the electrodes, the photo-thermoelectric effect would create a voltage between the junction and the electrode which is at a lower temperature. If the Seebeck coefficient is the same on both sides of the junction, the voltage on the two sides would be the same magnitude with opposite directions, resulting in zero net voltage between the electrodes. If the Seebeck coefficients are different, the mismatch would develop a net voltage difference between the two electrodes and thus generate photocurrent (Fig. S4b, Supplementary data). This is the photo-thermoelectric effect.

In order to elucidate the origin of the photocurrent, we carried out theoretical calculations of the gate voltage dependences of the photocurrent at the ABA/ABC junction (Fig. 4a) and BLG/SLG junction (Fig. S5a, Supplementary data). The number of electrons was converted into the gate voltage for comparison with the experimental results [31,44,45]. In the following discussion, a net electron current from the ABA side to the ABC side through the junction is defined as the positive current direction. In the ABA/ABC photodetector, a Fermi energy difference would lead to motion of electrons and holes as explained in Fig. S4 (Supplementary data). A positive $\Delta E_F$ would correspond to an electron moving from the ABC side to the ABA side, and hence a negative current. Therefore, we compare $-\Delta E_F$ value with the experimental photocurrent data in Fig. 4b. On the other hand, a positive Seebeck coefficient difference would mean that the voltage of the electrode on the ABA side is higher than on the ABC side, and hence a positive current flows. Therefore, we compare $\Delta S$ with the experimental photocurrent data in Fig. 4c.

The green curves in Figs. 4b and c are the calculated $-\Delta E_F$ and $\Delta S$ as functions of the back-gate voltage with respect to CNP. A small difference (~ 9 meV) in the work functions of ABA and ABC TLG is ignored. It is obvious that the calculated results cannot explain the experimental data. In this calculation, it was assumed that the offset between the Dirac points of ABA and ABC TLG in the ABA/ABC junction is the same as in the case when the two materials are separated. In other words, it was assumed that forming the junction does not affect the band alignment between ABA and ABC TLG. However, as it is well known from semiconductor heterostructures [46], the details of the interface play a critical role in determining the band lineups. Since there should be a sizeable strain at the ABA/ABC junction due to the lattice mismatch between the third A- and C-layers [47,48], a modification of the band lineup is naturally expected. Because such a strain is localized to within 7-11 nm of the junction [47,48], it is difficult to measure directly by Raman spectroscopy, for example. We therefore used the possible band lineup modifications as fitting parameters to fit the calculations to the experimental data. The blue curve in Fig. 4c was a best fit to the data, obtained by shifting the Dirac point energy of ABA TLG by –0.131 eV and that of ABC TLG by +0.005 eV. These shifts correspond to the CNP shifts of –82 V and +1V, respectively (Fig. S6c, Supplementary data). It should be noted that these shifts of the bands are in addition to the back-gate-induced shift of the band alignment due to different densities of states of ABA and ABC TLG. The blue curve for ΔS in Fig. 4c reproduces all the important features of the experimental data in Fig. 4a, whereas the blue curve for in Fig. 4b is qualitatively different from the experimental data. For comparison, we repeated the calculations with opposite movement of the bands, i.e., +0.131 eV and –0.005 eV shifts for Dirac point energies of ABA and ABC TLG, respectively. It is obvious that the calculated trends are completely inconsistent with the experimental data. From this exercise, we conclude that the Seebeck coefficient difference, i.e., the photo-thermoelectric effect is the dominant mechanism for the photocurrent in the ABA/ABC junction. As a comparison, we also analyzed the photocurrent in the BLG/SLG junction device (Fig. S5, Supplementary data). Again, the Seebeck coefficient difference seems to be the dominant mechanism for the photocurrent.

Fig. 4d summarizes the band alignment of lateral ABA/ABC junction. When a junction is formed, the Dirac point energies of the constituent materials shift with respect to their vacuum values, resulting in built-in band offsets. Owing to these built-in band offsets, a lateral potential difference is naturally created even without external back-gate bias or at the CNP. These offsets significantly modify the back-gate dependence of the Fermi energy difference and the Seebeck coefficient difference. Without assuming these offsets, the measured photocurrent data cannot be explained by the theoretical calculations.

Because of the increased optical absorption of TLG, the responsivity of the ABA/ABC TLG photodetectors should be larger than that of metal-SLG devices. Our ABA/ABC TLG device has the maximum responsivity of 0.18 A/W. These responsivity values are much larger than that of any other pure graphene photodetectors reported to date [17].

## 4. Conclusions

In summary, we measured the gate voltage dependence of the photocurrent in TLG photodetectors with both ABA- and ABC-stacked domains. The photocurrent is generated at the lateral junction between ABA- and ABC-stacked domains. By comparing the experimental data with theoretical calculations, we found that there is a built-in band offset between the two domains, and the dominant mechanism of the photocurrent is the photo-thermoelectric effect due to the Seebeck coefficient difference.

Acknowledgments

This work was supported by the National Research Foundation of Korea grants funded by the Ministry of Science, ICT and Future Planning of Korea (No. 2011-0017605 and 2012R1A2A2A01045496) and grants from the Center for Advanced Soft Electronics under the Global Frontier Research Program of the MSIP (No. 2011-0031630 and No. 2011-0031640).

Computations were supported by the CAC of KIAS. M.K. is supported by the TJ Park Science Fellowship of POSCO TJ Park Foundation.

**Appendix A. Supplementary data**

Supplementary data associated with this article can be found, in the online version.

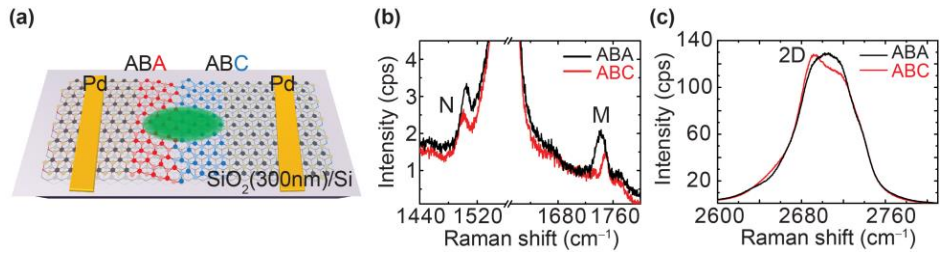

**Fig. 1.** (a) Schematic of a graphene photodetector with a lateral junction between ABA- and ABC-stacked domains in TLG (ABA/ABC photodetector). (b) Raman N, M and (c) 2D bands from ABA (black) and ABC (red) stacked regions of TLG.

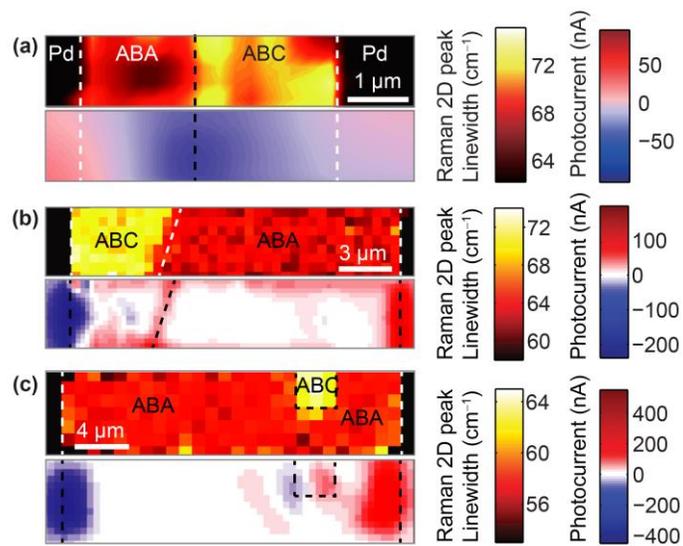

**Fig. 2.** Comparison of Raman 2D peak linewidth (top) and photocurrent (bottom) images of 3 different ABA/ABC photodetectors. Photocurrent is observed at the junctions between ABA- and ABC-stacked domains in all 3 devices.

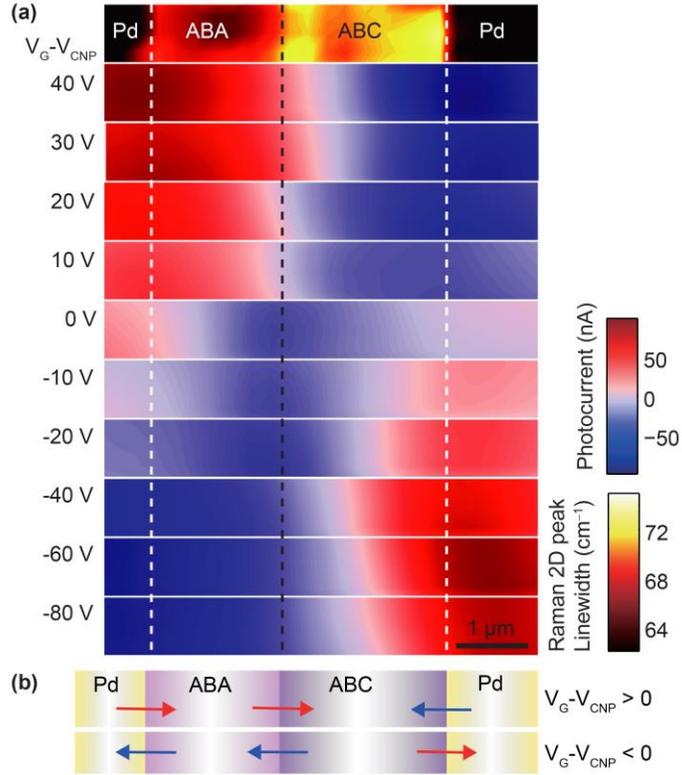

**Fig. 3.** (a) Raman 2D peak linewidth image (top) and photocurrent images as a function of gate voltage with respect to CNP ($V_G$-$V_{CNP}$). In the Raman image, the red (yellow) region in the Raman image indicates ABA- (ABC-) stacked TLG, whereas black regions indicate the electrodes (Pd). $V_G$ is the applied gate voltage and $V_{CNP}$ is the CNP voltage of the graphene channel. The red (blue) regions in the photocurrent images correspond to electron motion to the right (left). The black dashed line indicates the boundary between the ABA- and ABC-stacked domains. (b) Schematic diagram of photocurrent directions in ABA/ABC photodetector.

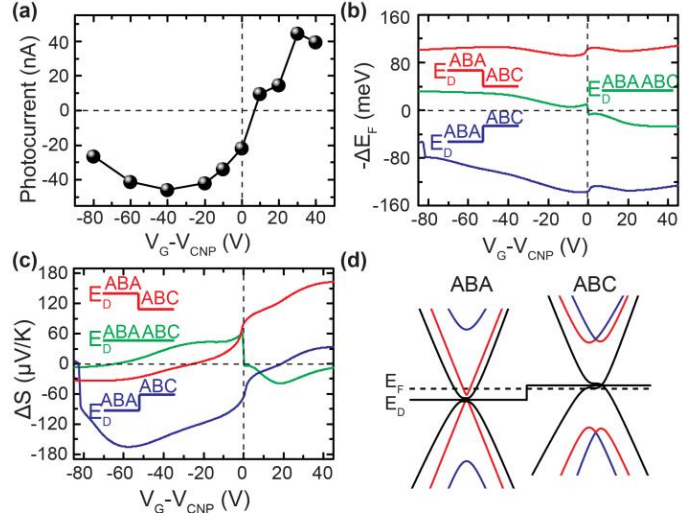

**Fig. 4.** Comparison of measured photocurrent and calculated Fermi energy and Seebeck coefficient difference. (a) Measured photocurrent as a function of gate voltage with respect to CNP ($V_G$-$V_{CNP}$) at ABA/ABC junction. (b) Calculated Fermi energy difference at ABA/ABC junction. (c) Calculated Seebeck coefficient difference at ABA/ABC junction. Green curves correspond to zero band offset, whereas red and blue curves correspond to band offsets as indicated. (d) Band alignment of the ABA/ABC junction in TLG for $V_G$=$V_{CNP}$.